\documentclass[9pt,conference]{IEEEtran}
\IEEEoverridecommandlockouts

\usepackage{setspace,tabularx}
\usepackage[tight,footnotesize]{subfigure}
\usepackage{times}
\usepackage{algorithm}
\usepackage[noend]{algorithmic}
\usepackage{amsfonts}
\usepackage{amsmath}
\usepackage{amssymb}
\usepackage{array}
\usepackage{boxedminipage}
\usepackage{caption}
\usepackage{cite}
\usepackage{color}
\usepackage{floatflt}
\usepackage{float}
\usepackage{graphics}
\usepackage{flushend}

\usepackage{graphicx}
\usepackage{indentfirst}
\usepackage{latexsym}
\usepackage{hyperref}
\usepackage{pifont}
\hypersetup{colorlinks=true,linkcolor=black,filecolor=black,urlcolor=black,citecolor=black}
\usepackage{multirow}
\usepackage{picinpar}
\usepackage{placeins}
\usepackage{psfrag}
\usepackage{soul}
\usepackage{subfigure}
\usepackage{theorem}
\usepackage[para]{threeparttable} 
\usepackage{wrapfig}
\usepackage{mathtools}
\usepackage{romannum}
\usepackage{graphicx}
\usepackage{enumitem}
\usepackage{colortbl}
\usepackage{amsmath,graphicx}
\usepackage{amsmath,amssymb,amsfonts}
\usepackage{bm}
\usepackage{graphicx}
\usepackage{textcomp}
\usepackage{xcolor}
\usepackage{mathrsfs}
\usepackage{pifont}
\newcommand\mynuma[1]{\ifcase#1 \or \ding{172}\or \ding{173}\or
  \ding{174}\or \ding{175}\or \ding{176}\or \ding{177}%
  \or \ding{178}\or \ding{179}\or \ding{180}\or \ding{181}\else *\fi\relax}

\theoremstyle{break}

\newcommand{\ignore}[1]{ }

\newcommand{\figlbl}[1]{\label{fig.{#1}}}
\newcommand{\figref}[1]{Fig.~\ref{fig.{#1}}}

\newcommand{\eqnlbl}[1]{\label{eqn:{#1}}}
\newcommand{\eqnref}[1]{Equation~(\ref{eqn:{#1}})}

\newcommand{\beq}{\begin{equation}}
\newcommand{\eeq}{\end{equation}}

\graphicspath{{../fig/}}
\begin{document}

\title{KyberMat: Efficient Accelerator for Matrix-Vector Polynomial Multiplication in CRYSTALS-Kyber Scheme via NTT and Polyphase Decomposition}

\author{Weihang Tan$^\star$, Yingjie Lao$^\dag$, and~Keshab K. Parhi$^\star$\\
$^\star$Department of Electrical and Computer Engineering, University of Minnesota, Minneapolis, MN 55455, USA\\
$^\dag$Department of Electrical and Computer Engineering, Clemson University, Clemson, SC 29634, USA\\
wtan@umn.edu, ylao@clemson.edu, parhi@umn.edu}

\maketitle

\begin{abstract}
CRYSTAL-Kyber (Kyber) is one of the post-quantum cryptography (PQC) key-encapsulation mechanism (KEM) schemes selected during the standardization process. This paper addresses optimization for Kyber architecture with respect to latency and throughput constraints. 
Specifically, matrix-vector multiplication and number theoretic transform (NTT)-based polynomial multiplication are critical operations and bottlenecks that require optimization. To address this challenge, we propose an algorithm and hardware co-design approach to systematically optimize matrix-vector multiplication and NTT-based polynomial multiplication by employing a novel {\em sub-structure sharing} technique in order to reduce computational complexity, i.e., the number of modular multiplications and modular additions/subtractions consumed. The sub-structure sharing approach is inspired by prior fast parallel approaches based on polyphase decomposition. The proposed efficient feed-forward architecture achieves high speed, low latency, and full utilization of all hardware components, which can significantly enhance the overall efficiency of the Kyber scheme. The FPGA implementation results show that our proposed design, using the fast two-parallel structure, leads to an approximate reduction of $90\%$ in execution time ($\mu s$), along with a $66\times$ improvement in throughput performance. 
\end{abstract}


\section{Introduction}

As part of the post-quantum cryptography (PQC) initiative, the NIST has identified and chosen the CRYSTALS-Kyber (Kyber) scheme as one of the recommended public-key encryption (PKE) and key-encapsulation mechanism (KEM) algorithm in 2022~\cite{bos2018crystals}.

Kyber is derived from the learning with errors (LWE) problem~\cite{regev2009lattices} that belongs to lattice-based cryptography. However, unlike other lattice-based cryptography schemes, the computational problem utilized in Kyber is module-learning with errors (M-LWE)~\cite{langlois2015worst} which requires matrix-vector and vector-vector polynomial (modular) multiplications. As the entries in the matrices and vectors are polynomials over the ring, all the polynomials are converted to their number theoretic transform (NTT)-domain representation to reduce the complexity when performing entry-entry multiplication. In addition, the latest Kyber scheme employs a special parameter setting that requires polyphase decomposition before performing the NTT-based polynomial multiplication, which results in a more complicated implementation~\cite{bos2018crystals}.

In fact, the integration of polyphase decomposition, fast filtering, NTT-based polynomial multiplication, sub-structure sharing, and matrix-vector polynomial multiplication in Kyber presents notable implementation and scheduling challenges not only for the algorithm but also for the hardware design. This paper presents a novel approach focused on co-designing hardware and algorithm for matrix-vector polynomial multiplication and NTT-based polynomial multiplication in Kyber. We propose a novel algorithm that leverages the sub-structure sharing technique~\cite{potkonjak1996multiple,parhi2007vlsi} for matrix-vector polynomial multiplication in the NTT-domain.

Based on the algorithmic optimization, an efficient hardware architecture design, \textbf{KyberMat}, for \textbf{Kyber} \textbf{mat}rix-vector polynomial multiplication using the NTT algorithm is presented. Due to the large data size in the Kyber, it becomes imperative for hardware architectures to exhibit fast data processing, efficient communication, and minimize data movement to memory. Consequently, the development of a high-throughput hardware implementation becomes crucial in order to enable the swift execution of computations and handle greater number of data sequences within a given accelerator. In addition, the proposed KyberMat accelerator uses feed-forward architecture with only one direction from input to output and is pipelined through different stages to ensure a short critical path. KyberMat accelerator achieves a high-speed, real-time, and high-throughput performance.

The contributions of this paper are summarized as follows:
\begin{itemize}
    \item We point out the connection between fast parallel finite impulse response (FIR) filter \cite{parhi2007vlsi,parker1997low,cheng2004hardware} and point-wise multiplication of polynomials in NTT-domain. This enables us to use higher-level parallelism, such as four- or eight-parallel, and different types of fast FIR filters. For example, prior work was limited to only specific two-parallel FIR structures in the context of a single polynomial modular multiplication~\cite{zhou2019preprocess,xing2021compact,zhu2021ntt}, as opposed to matrix-vector multiplication of polynomials.
    \item This paper presents {\em novel sub-structure sharing} \cite{potkonjak1996multiple,parhi2007vlsi} approaches for point-wise multiplication in matrix-vector polynomial multiplication  based on original-form and transpose-form fast FIR filters. The use of sub-structure sharing is the {\em key} to reduce the number of modular multiplications and additions.
    
    \item We present a novel and efficient algorithm for the matrix-vector polynomial multiplication for Kyber, which reduces the number of modular multiplications and additions required, compared to previous optimizations. To the best of our knowledge, this work is the first to systematically explore optimizations for matrix-vector multiplication in the NTT-domain for the Kyber scheme. 
    
    \item Furthermore, the parallelism of the architecture can be arbitrary; this will lower the latency and increase throughput at the expense of an increase in hardware. These architectures are ideal for cloud computing.

    \item Our experimental results demonstrate that the proposed KyberMat significantly enhances both execution time (measured in $\mu s$) and throughput performance over existing state-of-the-art designs.

\end{itemize}

The rest of this paper is structured as follows. Section~\ref{math} provides a brief overview of the Kyber scheme along with the related hardware architectures and algorithms in previous works. Section~\ref{sec_filter} presents the insight into the relationship between parallel FIR filter structure and NTT-based polynomial multiplication using polyphase decomposition. 
Section~\ref{algm_two} describes the proposed novel and efficient algorithm-hardware co-optimized KyberMat architecture. 
Section~\ref{result} presents the performance analysis of the proposed architecture, with detailed comparisons to previous works. Finally, Section~\ref{conclusion} concludes the paper. 

\section{Background}\label{math}
\subsection{Notation and parameter space}
In this paper, the single polynomial over the ring $R_q = \mathbb{Z}_{q}/(x^n+1)$ is denoted as $a(x)$. The bold symbols represent the polynomial vector,  $\bm{a} \in R^{k }_q$ and polynomial matrix  $\bm{A} \in R^{k \times k}_q$, whose entries are polynomials.  The notations $\bm{a}^T$ and $\hat{\bm{a}}$ denote the transpose of the matrix (or vector) and the NTT-domain representation of the variable, respectively, and $\circ$ symbol represents point-wise multiplication between two polynomials.

\subsection{Kyber scheme}
The Kyber scheme is a secure KEM that is indistinguishable under chosen-ciphertext attack (IND-CCA) and consists of three algorithms: key generation (KeyGen), encapsulation (Encaps), and decapsulation (Decaps)~\cite{bos2018crystals}.  It is primarily described as an indistinguishable under chosen-plaintext attack (IND-CPA) security public-key encryption (PKE) scheme, which can be further transformed into the IND-CCA secure KEM using the Fujisaki-Okamoto transform~\cite{fujisaki1999secure}.

Kyber provides three different security levels, i.e., Kyber-512, Kyber-768, and Kyber-1024, to satisfy NIST security levels $1$, $3$, and $5$, respectively. To scale the security level for the Kyber scheme, we only require to change the module dimension $k$ in $k = 2$, $3$, and $4$ with multiple fixed length-$n$ polynomials over the ring $R_q$. 

The central component of the Kyber scheme is the M-LWE sample, which requires computations over vector and matrix~\cite{bos2018crystals,ravi2022side}. Specifically, the Encaps algorithm generates two M-LWE samples $\bm{u} \in R_q^{k}$. For example,  
\begin{align}
    \bm{u} &= \bm{A}^T \bm{r} + \bm{e}_1, \eqnlbl{eqn_matrix}
\end{align}
where $\bm{A}\in R_q^{k\times k}$ and $\bm{r}\in R_q^k$ are the random matrix and vector, respectively. $\bm{e}_1 \in R_q^k $ is the noisy vector, sampled from the centered binomial distribution (CBD)~\cite{bos2018crystals}.

\subsection{Matrix-vector polynomial multiplication in Kyber scheme}
The operations required to perform on M-LWE samplers involve polynomial-based computations on matrices and vectors (module). The core operations and bottlenecks are matrix-vector and vector-vector polynomial multiplications since they involve polynomial modular multiplication and polynomial modular addition. The Kyber scheme incorporates the NTT-domain representation into its definition to reduce the computational complexity of polynomial modular multiplication. In particular, the random matrix $\bm{A}$ is naturally sampled in the NTT-domain as $\hat{\bm{A}}$, and the keys are also stored in the NTT-domain. 

To efficiently perform entry-entry multiplications in \eqnref{eqn_matrix}, NTT-based polynomial multiplication is used. 
This operation requires an NTT computation for the random vector $\bm{r}$, which is represented as $\hat{\bm{r}} = \text{NTT}(\bm{r})$. 

In general, the process of NTT-based polynomial multiplication involves converting the polynomials to their corresponding NTT-domain representations. These representations enable point-wise multiplication to generate the NTT-domain polynomial. The resulting polynomial is then transformed back to the original algebraic domain using an inverse NTT (iNTT) computation to obtain the polynomial product~\cite{lyubashevsky2008swifft}.

By using the NTT-domain representation, \eqnref{eqn_matrix} can be re-represented as 
\begin{align}
    \bm{u} &= \text{iNTT} (\hat{\bm{A}}^T \cdot \text{NTT}(\bm{r})) + \bm{e}_1. \eqnlbl{eq_ntt_u} 
\end{align}

The NTT-domain matrix-vector polynomial multiplication in \eqnref{eq_ntt_u} plays a critical role in the Kyber scheme due to its dominance with respect to the number of modular (integer) multiplications. Hence, optimizing these computationally intensive operations in hardware can significantly improve the performance of the Kyber scheme.

\subsection{Prior optimizations for Kyber scheme}\label{priorwork}
The latest version of the Kyber scheme chooses a new prime $q=3329$, which does not satisfy $q \equiv 1 \mod 2n$ when $n = 256$.  As a result, the NTT-based polynomial multiplication requires a polyphase decomposition, where the NTT computations rely on 128-point and a subsequent complex point-wise multiplication as presented in \cite{bos2018crystals} and 1PtNTT algorithm detailed in~\cite{zhou2019preprocess}. 
The prior hardware accelerations for Kyber~\cite{xing2021compact,bisheh2021instruction,aikata2022kali,hu2022ac} apply the polyphase decomposition before NTT computation, which then requires several rounds of 128-point NTT computation.

The prior work presented in~\cite{xing2021compact} and 1IPtNTT algorithm described in~\cite{zhu2021ntt} exploit fast convolution concepts to reduce the number of modular multiplications during point-wise multiplication, which is similar to the original fast filtering algorithm~\cite{parhi2007vlsi}. This method is subsequently adopted in later designs as in~\cite{bisheh2021instruction,aikata2022kali}.
However, these previous studies exclusively focus on optimizing NTT-based polynomial multiplication utilizing polyphase decomposition for a two-parallel design only. They do not concurrently take into account the optimization of matrix-vector polynomial multiplication in the Kyber scheme, thereby leaving unexplored design space that can further reduce computational complexity. 

The paper points out the connection between fast FIR filter and point-wise multiplication in NTT-domain. This allows the use of higher-level parallelism such as four-parallel or eight-parallel in polynomial multiplication. Then the paper considers matrix-vector polynomial multiplication and proposes novel sub-structure sharing to further reduce the number of multiplications for point-wise multiplication. Sub-structure sharing has been used both at the algorithm level \cite{lucke94,parhi2007vlsi} and at the hardware level \cite{potkonjak1996multiple,parhi2007vlsi}. Sub-structure sharing leads to significant reduction in the complexity of the proposed architectures.

\begin{figure*}[htbp]
\centering
\resizebox{1.0\textwidth}{!}{
\includegraphics{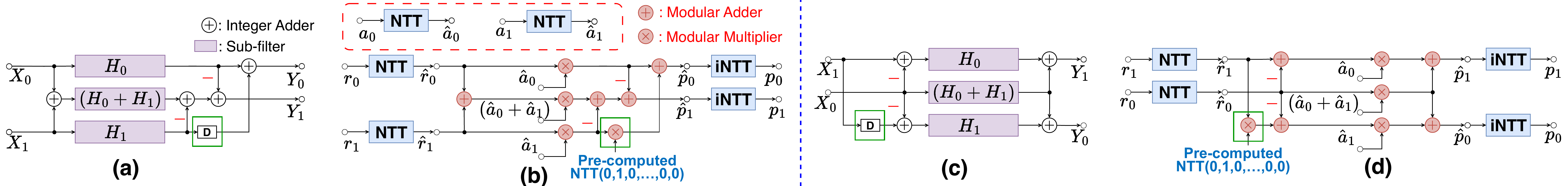}}
\caption{\small{Data-flow graph for two-parallel fast filtering structure and NTT-based polynomial multiplication using polyphase decomposition. (a) Original two-parallel fast filtering structure. (b) NTT-based polynomial multiplication using original parallel fast filtering structure. (c) Transposed two-parallel fast filtering structure. (d) NTT-based polynomial multiplication using transposed parallel fast filtering structure.}}
\figlbl{relationship_fir_ntt}
\vspace{-1.5em}
\end{figure*}

\section{Relationship Between Parallel FIR Filter Structure and Polynomial Modular Multiplication using NTT and Polyphase Decomposition}\label{sec_filter}

The FIR filter is one of the important elements in digital signal processing.  
The FIR filter is also applied to perform the convolution on a digital signal with a finite number of taps. Efficient hardware and software implementations of the FIR filter have been widely studied~\cite{oppenheim2009discrete,yuan2020high,denkinger2022vwr2a,cheng2020fast}. In particular, the fast filtering algorithm and its structure (i.e., fast filtering structure) have been used to increase the parallelism and reduce complexity, ultimately improving throughput performance~\cite{parhi2007vlsi}.  Fast filtering structures, as represented in \figref{relationship_fir_ntt}(a) and \figref{relationship_fir_ntt}(c), exhibit the same computational complexity~\cite{parhi2007vlsi}. However, they differ in the data flow. \figref{relationship_fir_ntt}(a) displays the original fast filtering structure, while \figref{relationship_fir_ntt}(c) demonstrates its equivalent transposed structure.

\begin{figure*}[htbp]
\centering
\resizebox{0.98\textwidth}{!}{
\includegraphics{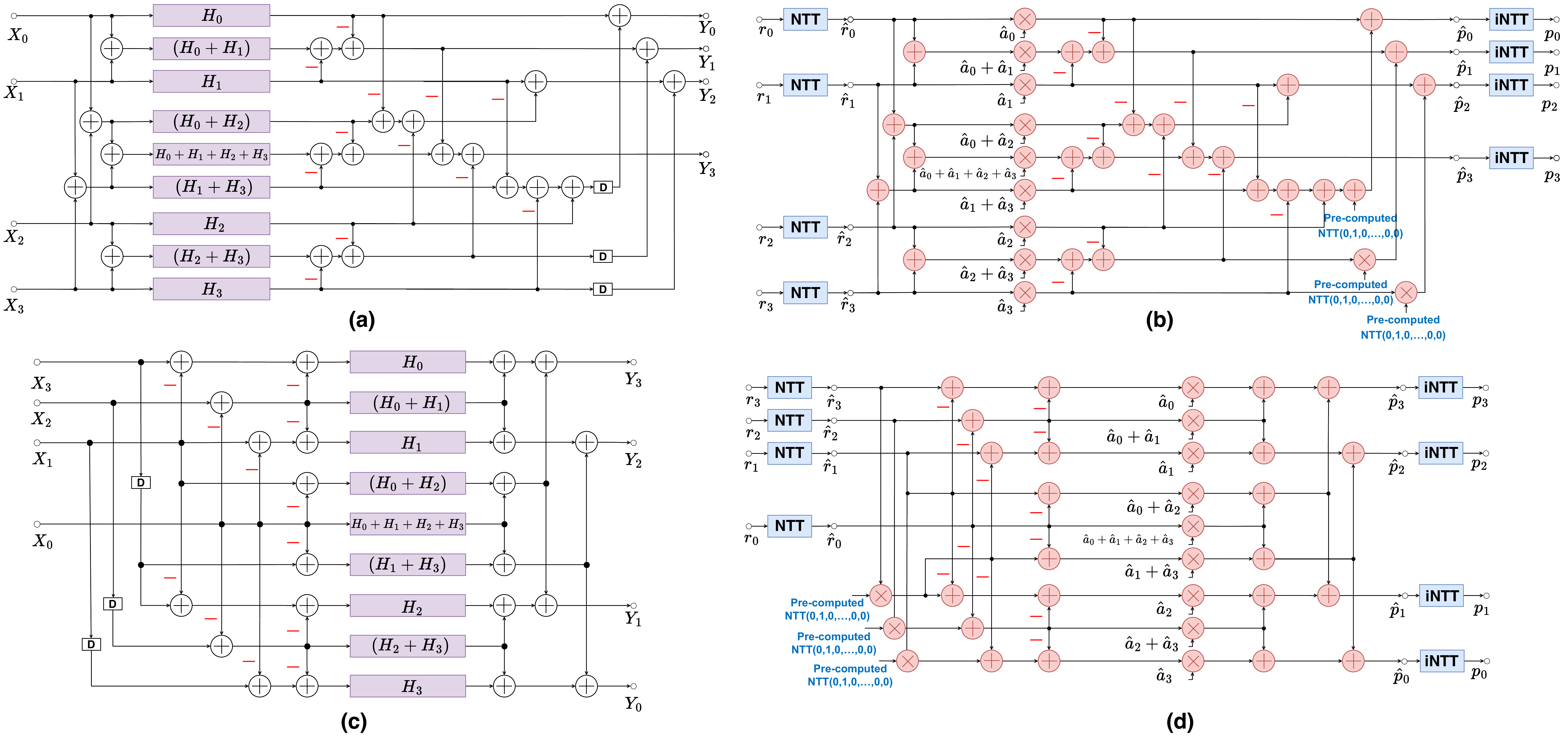}}
\caption{\small{Data-flow graph for four-parallel fast filtering structure and NTT-based polynomial multiplication using polyphase decomposition. (a) Original four-parallel fast filtering structure. (b) NTT-based multiplication using original parallel fast filtering structure. (c) Transposed four-parallel fast filtering structure. (d) NTT-based multiplication using transposed parallel fast filtering structure.}}
\figlbl{fast_four_structure}
\end{figure*}
The process of computing the fast filtering algorithm $Y(z) = H(z)X(z)$ first involves the polyphase decomposition~\cite{oppenheim2009discrete}. The input sequence $ {x[0],x[1],x[2],\cdots}$ is initially represented as $X(z) = {x[0] + x[1]z^{-1} + x[2]z^{-2} +\cdots}$ in the $z$-domain, which then executes the polyphase decomposition $X(z) = X_0(z^2) + X_1(z^2)\cdot z^{-1}$, where $X_0(z^2)$ and $X_1(z^2)$ are $Z$-transforms of the even indexed-terms ($x[2l]$) and odd indexed-terms ($x[2l+1]$), respectively. The filter coefficients $H(z)$ undergo a similar polyphase decomposition to obtain $H_0(z^2)$ and $H_1(z^2)$. 

The outputs of the fast filtering algorithm are expressed as:
\begin{align}
    Y_0(z^2) &= X_0(z^2)H_0(z^2) + z^{-2}X_1(z^2)H_1(z^2) \eqnlbl{eq_firy_0}\\
    Y_1(z^2) &= X_0(z^2)H_1(z^2) + X_1(z^2)H_0(z^2) \nonumber \\ 
             &= \left (H_0(z^2) + H_1(z^2) \right )\left (X_0(z^2) + X_1(z^2)\right ) \nonumber\\
             &- X_0(z^2)H_0(z^2) - X_1(z^2)H_1(z^2), \eqnlbl{eq_firy_1}
\end{align}
where $Y(z) = Y_0(z^2) + Y_1(z^2)\cdot z^{-1}$.

Such operation involves three length-$\frac{n}{2}$ point-wise multiplications and five length-$\frac{n}{2}$ point-wise additions/subtractions, as illustrated in \figref{relationship_fir_ntt}(a) and \figref{relationship_fir_ntt}(c). The delay element boxed in green plays the role of multiplication with $z^{-2}$ in a two-parallel architecture. The fast filter approach has been exploited to reduce the number of operations in the polynomial modular multiplication in the time domain~\cite{tan2023high}. In this context, the delay element in the fast filter is equivalent to multiplication by $x^2$.

The focus of this paper is the use of fast filter approaches to reduce the number of multiplications in the frequency domain. Here the polynomial modular multiplication is described in the frequency domain first. For a general polynomial modular multiplication $p(x) = r(x)\cdot a(x) \mod (x^n+1)$, its NTT representation is defined as
\begin{align}
    p(x) &= \text{iNTT}\left (\text{NTT}(r(x))\circ \text{NTT}(a(x))\right ) \nonumber\\
    &= \text{iNTT}\left (\hat{r}(x) \circ \hat{a}(x)\right ).
\end{align}

By leveraging the polyphase decomposition and fast filtering algorithm for the NTT-based polynomial multiplication, the \eqnref{eq_firy_0} and \eqnref{eq_firy_1} can be expressed as
\begin{equation}
    p_0(x^2) = \text{iNTT}\big(\hat{r}_0(x^2)\circ\hat{a}_0(x^2) + x^{2}\cdot\hat{r}_1(x^2)\circ\hat{a}_1(x^2)\big)  \eqnlbl{eq_ntty_0}
\end{equation}
\begin{align}
    p_1(x^2) = \text{iNTT}  \big(&\hat{r}_0(x^2)\circ\hat{a}_1(x^2) + \hat{r}_1(x^2)\circ\hat{a}_0(x^2) \big) \nonumber \\ 
             = \text{iNTT} \Big(&  \big(\hat{r}_0(x^2)+\hat{r}_1(x^2)  \big) \circ  \big(\hat{a}_0(x^2)+\hat{a}_1(x^2)  \big) \nonumber \\
              &- \hat{r}_0(x^2)\circ\hat{a}_0(x^2) - \hat{r}_1(x^2)\circ\hat{a}_1(x^2)\Big) \eqnlbl{eq_ntty_1},
\end{align}
where $\hat{r}_0(x^2)$, $\hat{r}_1(x^2)$, $\hat{a}_0(x^2)$, and $\hat{a}_1(x^2)$ represent the NTT of the input polynomials after polyphase decomposition, and $p(x) =p_0(x^2) + p_1(x^2) \cdot x$.

To apply the fast filtering algorithm to NTT-based polynomial multiplication, this work transforms \figref{relationship_fir_ntt}(a) and \figref{relationship_fir_ntt}(c), into NTT-based structures in the frequency domain, as shown in \figref{relationship_fir_ntt}(b) and \figref{relationship_fir_ntt}(d). However, directly utilizing the delay element in the time domain to represent multiplication by $x^2$ is not feasible in the NTT-domain. Instead, a point-wise multiplication with a pre-computed constant set, $\text{NTT}(x^2)$, of length $\frac{n}{2}$ is utilized. It may be noted that the structure in \figref{relationship_fir_ntt}(c) is equivalent to the structure referred to as 1IPtNTT algorithm in in~\cite{zhu2021ntt} and optimized algorithm in~\cite{xing2021compact}. Higher-level parallelism can also be used. For example, \figref{fast_four_structure}(a) and \figref{fast_four_structure}(c) present the fast four-parallel structures for efficient FIR filter design in~\cite{cheng2004hardware}. Note that equivalent fast structures based on sub-filters ($H_0 ~-~ H_1$) can also be used instead of ($H_0 ~+~ H_1$)~\cite{parhi2007vlsi}.

\begin{figure*}[htbp]
\centering
\resizebox{1\textwidth}{!}{
\includegraphics{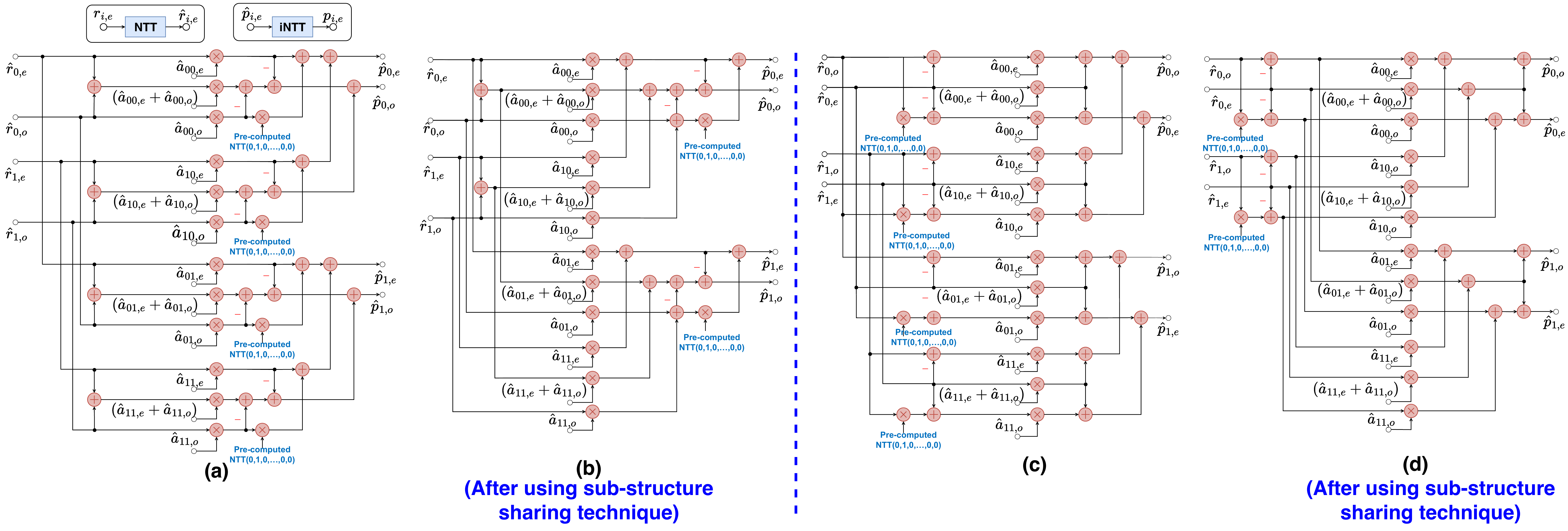}}
\caption{\small{Data-flow diagram illustrating the matrix-vector polynomial multiplication algorithm for the Kyber scheme using NTT and polyphase decomposition when $k=2$ using fast two-parallel structure (NTT/iNTT computations at the top are omitted for simplicity).  (a) Original form structure before our optimization. (b) Original form structure with sub-structure sharing. (c) Transposed form structure before our optimization. (d) Transposed form structure with sub-structure sharing.}}
\figlbl{sub_structure_sharing}
\vspace{-1em}
\end{figure*}

\begin{figure}[htbp]
\centering
\resizebox{0.5\textwidth}{!}{
\includegraphics{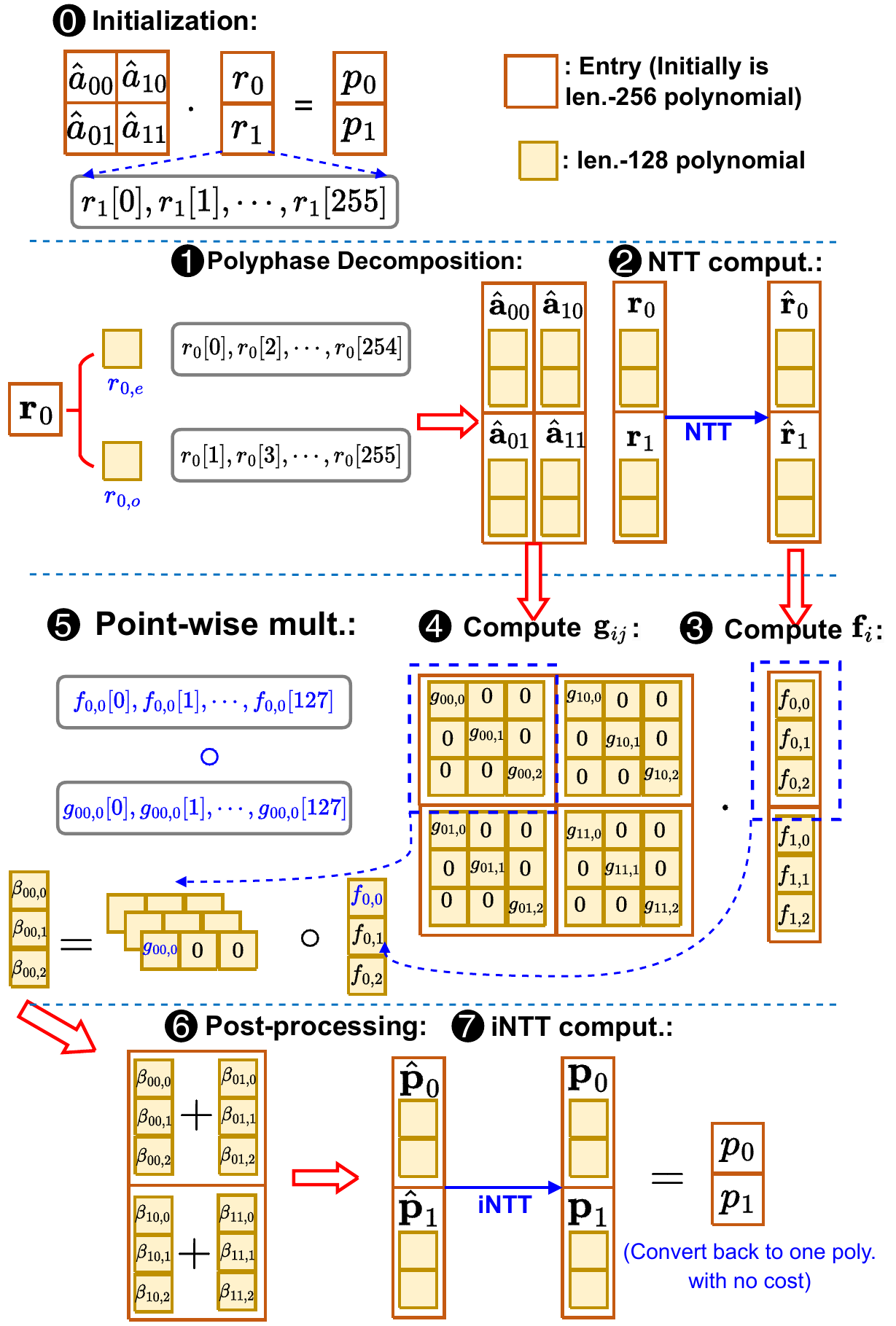}}
\caption{\small{Overview of our proposed efficient algorithm for KyberMat accelerator design when $k=2$.}}
\figlbl{matrix_vector_mult}
\end{figure}
\section{Algorithm-Hardware Co-Optimized KyberMat Architecture} \label{algm_two}
To employ the NTT algorithm and polyphase decomposition for polynomial modular multiplication for the Kyber scheme, the earlier studies utilized the traditional two-parallel FIR filter implementation for the matrix-vector polynomial multiplication in NTT-domain~\cite{zhou2019preprocess,bos2018crystals}. To minimize the number of modular multiplication in the point-wise multiplication, additional optimizations reduced the five length-$\frac{n}{2}$ point-wise multiplications down to four for each  entry-entry multiplication~\cite{xing2021compact,zhu2021ntt}. A data-flow graph, based on an example when $k=2$ from~\cite{xing2021compact}, and derived from \figref{relationship_fir_ntt}(b), is shown in \figref{sub_structure_sharing}(a). Its equivalent transposed structure is depicted in \figref{sub_structure_sharing}(c). However, all of these prior optimizations still necessitate executing length-$\frac{n}{2}$ point-wise multiplication with $\text{NTT}(x^2)$ for each entry-entry multiplication, leading to $k^2$ such operations in total for a single matrix-vector polynomial multiplication.

Different from these prior works,  this paper proposes the KyberMat architecture, a novel and efficient algorithm and hardware co-optimization for matrix-vector polynomial multiplication in the Kyber scheme. In this section, we first use the transposed structure (\figref{sub_structure_sharing}(c)) as a baseline example design to demonstrate our optimization by utilizing the sub-structure sharing technique to reduce computational complexity. Subsequently, we extend and generalize this optimization to the original structure (\figref{relationship_fir_ntt}(a)). We also show different fast filtering algorithms and structures that can be utilized to realize various benefits. Finally, a low-latency architecture design for KyberMat is presented. It is important to note that the use of {\em sub-structure sharing} in the fast NTT structures is the {\em key} to achieving hardware savings in the proposed KyberMat architecture. The sub-structure sharing is achieved in a natural way in the fast transpose structure and after applying {\em distributivity and associativity} in the original fast structure.

\subsection{Efficient algorithm of KyberMat using transposed two-parallel fast filtering structure}
The proposed algorithm for KyberMat to compute $\bm{p} = \bm{A}^T \bm{r} \in R^k_q$ is illustrated in Algorithm~\ref{algm_matrix_mult}, which consists of three stages: (i) pre-processing for the input matrix and vector (Lines 1-8), (ii) efficient point-wise multiplication in NTT-domain (Lines 9-11), and (iii) post-processing  (Lines 12-18). \figref{matrix_vector_mult} shows the overview and an example for our proposed algorithm when $k=2$.

\begin{algorithm}[htbp]
\caption{\textbf{Efficient Matrix-Vector Polynomial Multiplication for Kyber}}
\label{algm_matrix_mult}
\hspace*{\algorithmicindent}
\textbf{Input:}  $\hat{\bm{A}}^T$ and $\bm{r}$

\hspace*{\algorithmicindent} \textbf{Output:} $\bm{p} = \bm{A}^T \bm{r} \in R^k_q$

\begin{algorithmic}[1]

    \FOR{$i = 0$ \TO $k-1$ } {
                    \STATE $r_i(x) = r_{i,e}(x^2) + r_{i,o}(x^2)\cdot x$
                    \STATE $\hat{r}_{i,e}=\text{NTT}(r_{i,e}(x^2))$; $\hat{r}_{i,o} = \text{NTT}(r_{i,o}(x^2))$\\
                    \STATE $
                    f_{i,\{0,1,2\}}
                    = \{\hat{r}_{i,o} - \hat{r}_{i,e},\hat{r}_{i,e},\hat{r}_{i,o} \circ 
                    \text{NTT}(x^2) 
                    - \hat{r}_{i,e}$\}\\
                           
        }\ENDFOR
     \FOR{$i = 0$ \TO $k-1$}{
        \FOR{$j = 0$ \TO $k-1$}{
                    \STATE $\hat{a}_{ij} = \hat{a}_{ij,e} + \hat{a}_{ij,o}\cdot x$
                    \STATE  $g_{ij,\{0,1,2\}} = \{\hat{a}_{ij,e},\hat{a}_{ij,e}+\hat{a}_{ij,o},\hat{a}_{ij,o}\}$
        }\ENDFOR                    
    }\ENDFOR
    \FOR{$i = 0$ \TO $k-1$}{
        \FOR{$j = 0$ \TO $k-1$}{
            \STATE $\beta_{ij,\{0,1,2\}} = g_{ji,\{0,1,2\}} \circ f_{i ,\{0,1,2\}}$                    
        }\ENDFOR
    }\ENDFOR
    \FOR{$i = 0$ \TO $k-1$}{
        \FOR{$j = 0$ \TO $k-1$}{
            \STATE  $sum_{i,\{0,1,2\}} =sum_{i,\{0,1,2\}} + \beta_{ij,\{0,1,2\}}$   
        }\ENDFOR
    }\ENDFOR
    
    \FOR{$i = 0$ \TO $k-1$}{
        \STATE $\hat{p}_{i,e} = sum_{i,1} + sum_{i,2}$; $\hat{p}_{i,o}= sum_{i,1} + sum_{i,0} $\\
        \STATE $p_{i,e}(x^2)=\text{iNTT}(\hat{p}_{i,e})$; $p_{i,o}(x^2) = \text{iNTT}(\hat{p}_{i,o})$\\
        \STATE $p_i(x) = p_{i,e}(x^2) + p_{i,o}(x^2)\cdot x$
    }\ENDFOR

\end{algorithmic}
\end{algorithm}

The first step required in matrix-vector polynomial multiplication in \eqnref{eq_ntt_u} is the NTT computation for polynomial-based entries in vector $\bm{r}$. As required by Kyber, each polynomial inside the vector initially undergoes a polyphase decomposition. Note that after polyphase decomposition, each entry in the matrix or vector becomes a vector with two polynomials, i.e.,  $\bm{r}_i = [r_{i,e} (x^2),r_{i,o} (x^2)]^T$, for $i\in[0,k-1]$, as elaborated in \figref{matrix_vector_mult} Step \ding{182}. 

To perform the entry-entry multiplication in the NTT-domain, two 128-point NTT computations are required for each entry (\figref{matrix_vector_mult} Step \ding{183}). Since matrix $\hat{\bm{A}}^T$ is naturally in NTT representation after sampling, no NTT computation is required. Nevertheless, each entry $\hat{a}_{ij}(x)$ in the matrix has to perform the polyphase decomposition, i.e., $\bm{\hat{a}}_{ij} = [\hat{a}_{ij,e} (x^2),\hat{a}_{ij,o}(x^2) ]^T$,  $j \in[0,k-1]$, so all the coefficients in even indexed-terms and odd indexed-terms polynomials are aligned when executing the point-wise multiplication. 

As described in \figref{matrix_vector_mult} Step \ding{184} and outlined in Line 4 of Algorithm~\ref{algm_matrix_mult}, each vector $\bm{\hat{r}}_i$, $i\in[0,k-1]$ is transformed into a new vector $\bm{f}_i  = [f_{i,0},f_{i,1}, f_{i,2}] \in R^3_{n/2}$ with three length-$\frac{n}{2}$ polynomials. In a similar fashion, each $\bm{\hat{a}}_{ij}$, $i \in [0, k-1]$, $j \in [0, k-1]$ is redefined as $\bm{g}_{ij} = [g_{ij,0},g_{ij,1},g_{ij,2}]\in R^3_{n/2}$, as illustrated in Lines 5-8 in Algorithm~\ref{algm_matrix_mult} and \figref{matrix_vector_mult} Step \ding{185}.

After the pre-processing stage, a total of $3k^2$ point-wise multiplications are executed for the polynomials in $\bm{f}_{i}$ and $\bm{g}_{ji}$, $i,j, \in [0,-k]$. As a result, $3k^2$ intermediate products $\beta_{ij}$ are produced, as illustrated in \figref{matrix_vector_mult} Step \ding{186} (Lines 9-11 in Algorithm~\ref{algm_matrix_mult}). 
As $\bm{\hat{A}}$ is transposed before the matrix-vector polynomial multiplication in \eqnref{eq_ntt_u}, $\bm{f}_{i}$ is multiplied by $\bm{g}_{ji}$ instead of $\bm{g}_{ij}$. 

The post-processing stage, presented in Lines 12-16 in Algorithm~\ref{algm_matrix_mult} and \figref{matrix_vector_mult} Step \ding{187}, only requires computing the sum of $\beta_{ij}$ in each row by additions. Subsequently, these sums are combined to form $\hat{p}_{i,e}(x^2)$ and $\hat{p}_{i,o}(x^2)$, $i\in [0,k-1]$.

The corresponding data-flow graph for Algorithm~\ref{algm_matrix_mult} is shown in \figref{sub_structure_sharing}(d), which demonstrates a significant reduction in the required number of components compared to the data-flow graph depicted in \figref{sub_structure_sharing}(c) having the same functionality.
The optimization of our proposed algorithm relies on the sub-structure sharing technique, which can be explained as follows.

The prior literature requires multiplications with $\text{NTT}(x^2)$ in each entry-entry multiplication, thereby resulting in $k^2$ length-$\frac{n}{2}$ point-wise multiplications~\cite{xing2019efficient,bisheh2021instruction,aikata2022kali,li2022reconfigurable,hu2022ac}. However, this approach does not integrate steps across different levels. Further optimization could be achieved by minimizing the repetitive computations that are shared across different operational levels, such as matrix-vector multiplication and NTT-based polynomial multiplication using polyphase decomposition. The proposed optimized algorithm, however, employs a sub-structure sharing technique to reduce the number of point-wise multiplications with $\text{NTT}(x^2)$ from $k^2$ to $k$. Furthermore, this algorithm performs the summation of entry-entry products, $\beta_{ij}$,  prior to their combination back into two polynomials, to minimize the number of point-wise additions.

The matrix-vector polynomial multiplication requires the dot-product in each row of $\bm{\hat{A}}^T$ to multiply with the same $\bm{\hat{r}}$. This algorithm can reduce the total computational cost by increasing the reuse opportunities for the intermediate results achieved from the expensive operation. Since the operation of point-wise multiplication is expensive, this algorithm arranges the multiplications with $\text{NTT}(x^2)$ and the vector $\bm{\hat{r}}_i$ in the pre-processing stage by leveraging the \textit{transposition property} from the transposed two-parallel fast filtering structure for the sub-structure sharing. Therefore, $\bm{f}_i$, $i \in [0,k-1]$ containing the intermediate result from the expensive operation can then be shared by the entry-entry multiplication as illustrated in \figref{sub_structure_sharing}(d).

The data-flow graph shown in \figref{sub_structure_sharing}(a) can be optimized to reduce the number of modular multiplications and additions by applying the sub-structure sharing, i.e., the sub-structure NTT$(x^2)$, and exploiting {\em distributivity} property of multiplication and {\em associativity} property of add operations to utilize the sub-structure sharing technique in our proposed Algorithm~\ref{algm_matrix_mult}. These optimizations allow relocation of point-wise multiplication with $\text{NTT}(x^2)$ to occur after the summation of the intermediate results $\beta_{ij,2}$ rather than before. This optimization can be described by:
\begin{align}
    &\big(\beta_{00,0} + \text{NTT}(x^2)\beta_{00,2}\big)  + \big(\beta_{01,0} +\text{NTT}(x^2)\beta_{01,2}\big) \nonumber \\
    =& (\beta_{00,0} + \beta_{01,0}) +\text{NTT}(x^2)(\beta_{00,2}+\beta_{01,2}).
\end{align}
This reordering minimizes the total number of expensive point-wise multiplications.
The optimized data-flow graph is presented in~\figref{sub_structure_sharing}(b), and its computational complexity is same as that of \figref{sub_structure_sharing}(d). It is important to note that the sub-structure sharing is applied in different ways to the original and transpose structures; however, both designs have the same computational complexity after applying the sub-structure sharing technique. Both structures can be used interchangeably.

\subsection{KyberMat using various fast filtering structures and levels of parallelism}\label{sec_four_parallel}

As presented in Section~\ref{sec_filter}, matrix-vector polynomial multiplication using the NTT algorithm can be designed by exploiting various types of fast filter approaches that are well-known in the signal processing literature~\cite{cheng2004hardware,parhi2007vlsi,mou1991short,parker1997low}. The fast filter algorithms are non-unique. The transpose form of a fast filter structure is another equivalent fast filter. Higher-length parallel filters can be designed by either iterating shorter-length filters or by using iterated fast convolution algorithms followed by post-processing. The reader is referred to the textbook for a detailed discussion on this topic~\cite{parhi2007vlsi}. 

We utilize a fast four-parallel (filtering) structure ($L=4$) in~\cite{cheng2004hardware} as a case study to demonstrate complexity reduction of the four-parallel polynomial modular multiplication. 
The data-flow graph for matrix-vector polynomial multiplication, when $k=2$ using a fast four-parallel transposed structure and sub-structure sharing technique, is shown in \figref{four_ffa_trans}. This example examines our proposed optimization for enhanced parallelism and higher throughput architecture due to the expansion of the number of input and output data-path. Specifically, each polynomial $r_i(x)$ is decomposed into four polynomials of length-64 in the polyphase decomposition step: $ r_i(x) = r_{i,0}(x^4) + r_{i,1}(x^4) \cdot x+  r_{i,2}(x^4) \cdot x^2 + r_{i,3}(x^4) \cdot x^3$, denoted as $\bm{r}_i = [r_{i,0}(x^4), r_{i,1}(x^4), r_{i,2}(x^4), r_{i,3}(x^4)]^T$. Consequently, the 64-point NTT/iNTT computations and length-64 point-wise multiplications can be utilized. 

Note that the depicted graph only illustrates the components for computing $\hat{\bm{p}}_0$, as the structures for computing $\hat{\bm{p}}_0$ and $\hat{\bm{p}}_1$ are similar in terms of their point-wise multiplication and post-processing stages. 
Our proposed sub-structure sharing technique provides a notable advantage of reduced computational complexity. As the structure's data-paths increase, the complexity reduction is achieved by decreasing the number of required point-wise multiplications and point-wise additions in each data-path.

\begin{figure}[htbp]
\centering
\resizebox{0.48\textwidth}{!}{
\includegraphics{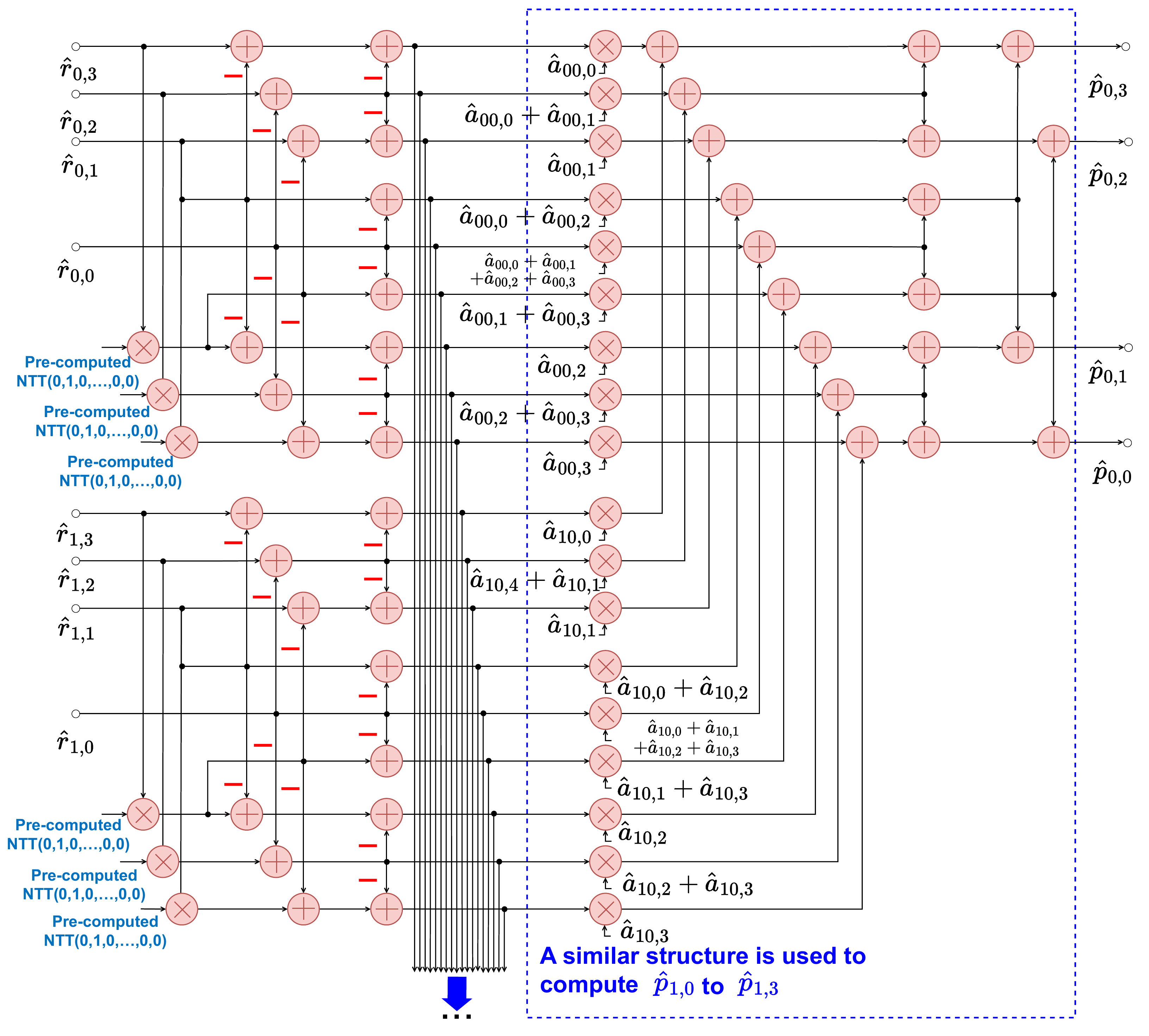}}
\caption{\small{Data-flow graph for matrix-vector polynomial multiplication when $k=2$ using fast four-parallel transposed structure  and sub-structure sharing technique. Components for computing $\hat{\bm{p}}_1$ are omitted.}}
\figlbl{four_ffa_trans}
\end{figure}

In addition to the reduction of computational complexity, employing a fast four-parallel structure in the hardware implementation of the algorithm can also reduce the latency of the system as the parallelism is increased. Since each component is responsible for only $\frac{n}{4}$ modular multiplications, the latency consumption is halved compared to the fast two-parallel structure.

Significantly, utilizing a fast eight-parallel structure framework ($L=8$) offers further enhancements to both throughput and latency performance. The application of our sub-structure sharing technique leads to a substantial reduction in computational complexity. For example, when employing the structure in \cite{cheng2004hardware}, the number of point-wise multiplications with $\text{NTT}(x^2)$ diminishes from $k^2(L-1)$ to $k(L-1)$.

\subsection{Efficient low-latency implementation for KyberMat} \label{archi}

This paper uses the data-flow graph in~\figref{sub_structure_sharing}(d) as an example to introduce the proposed low-latency design for KyberMat accelerator. The proposed low-latency design for KyberMat accelerator is illustrated in \figref{top_archi_high_speed}. The first building block is the NTT computation module that duplicates $2k$ 128-point NTT processors to convert all the polynomials in $\bm{r} = [\bm{r}_0,\bm{r}_1,\cdots,\bm{r}_{k-1}]$ to NTT-domain simultaneously, where $\bm{r}_i = [r_{i,e} (x^2),r_{i,o} (x^2)]^T$. The NTT/iNTT processors are instantiated by the optimized radix-2 multi-path commutator (R2MDC)-based architecture for NTT/iNTT computation~\cite{tan2023parentt,zhao2022high,nejatollahi2020exploring,hirner2023proteus} reconfigured for Kyber's parameter setting to satisfy the design criteria with real-time, multi-channels and feed-forward architecture. Besides, two input data-paths are used in each R2MDC-based architecture to increase the accelerator's throughput, similar to FFT architectures \cite{ayinala2011pipelined}. Each 128-point NTT processor structure consists of seven modular multipliers and fourteen modular adders/subtractors, resulting in  $14k$ modular multipliers and $28k$ modular adders/subtractors for the entire NTT computation module.

The next building block is the proposed novel matrix-vector polynomial multiplication in NTT-domain module, as shown in \figref{top_archi_high_speed}. This architecture can compute all the point-wise multiplications between the polynomials in $\bm{f}_{i}$ and $\bm{g}_{ji}$ simultaneously. As illustrated on the right-hand-side in \figref{top_archi_high_speed}, it maps each length-$\frac{n}{2}$ point-wise multiplication and addition in~\figref{sub_structure_sharing}(d) into two modular multipliers and two modular adders for upper and lower data-paths. As a result, $(6k^2+2k)$ modular multipliers and $(8k^2+2k)$ modular adders/subtractors are employed in  the matrix-vector polynomial multiplication in NTT-domain module. The hardware consumption in the iNTT computation module also employs $2k$ 128-point iNTT processors. As a result, the architecture requires $14k$ modular multipliers and $56k$ modular adders/subtractors in total for iNTT computation.

\begin{figure*}[htbp]
\centering
\resizebox{0.9\textwidth}{!}{
\includegraphics{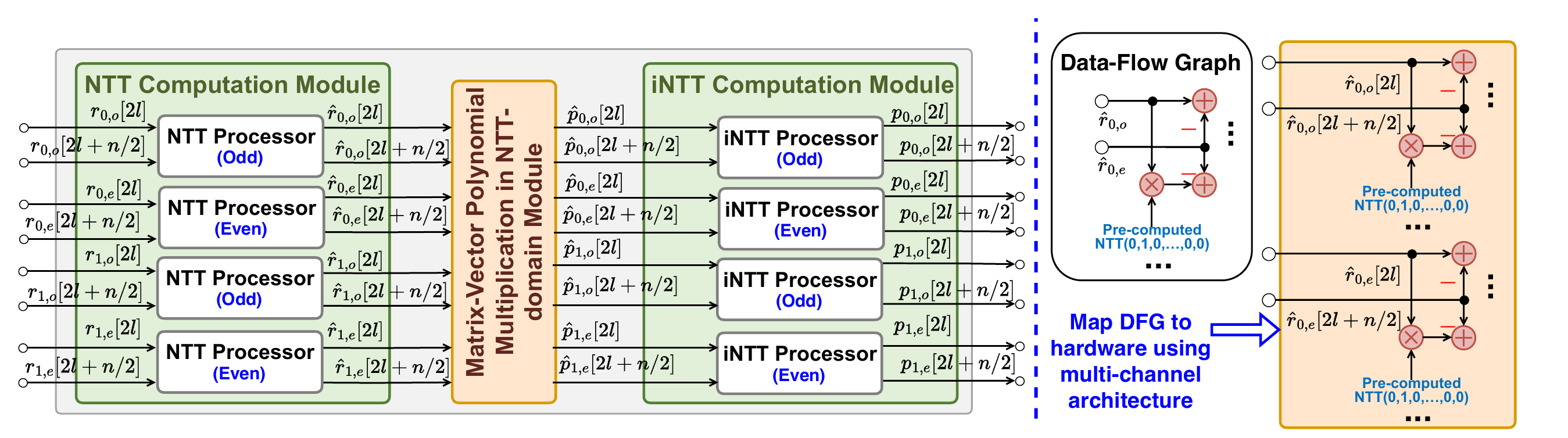}}
\caption{\small{Top-level architecture of low-latency design for KyberMat when $k=2$.}}
\figlbl{top_archi_high_speed}
\end{figure*}

The main advantage of the low-latency architecture design  for KyberMat is the significantly reduced clock cycle consumption and increased throughput. In contrast to previous works that require a large number of clock cycles for point-wise multiplication, the low-latency design parallelizes more modular multipliers in the data-path, reducing the latency in point-wise multiplication to only a few clock cycles utilized for pipelining.

\begin{table*}[htbp]
  \centering
  \caption{Performance of the proposed KyberMat accelerator design and prior works for Kyber-512 ($k=2$, $n=256$) in Artix-7 FPGA}\label{tb_mul}
\begin{tabular}{|c||c|c|c|c|c|c|}
\hline
Design  &LUTs (AT$^2$P $\times10^4$)  &FFs &DSPs (AT$^2$P $\times 10^2$)   &Freq.[MHz] &Cycles ($\mu s$) &TP[Gb/s]\\ \hline 
Xing~\cite{xing2021compact}  & 1737 (68.37) &1167 &2 (7.87)  &161  &3200 (19.84) &0.31\\  \hline 
Guo~\cite{guo2021efficient} & 1549 (16.27) &788 &4 (4.12)   &159  &1614 (10.15)  &0.61\\  \hline 
Bisheh~\cite{bisheh2021instruction} & 720 (121.33) &290 &6 (101.11)   &115  &4721 (41.05)  &0.15\\  \hline 
Bisheh (Parallel)~\cite{bisheh2021instruction} & 1474 (148.57) &580 &12 (121.12)   &115  &3654 (31.77)  &0.19\\  \hline 
Zhao~\cite{zhao2022high}  & 25674 (22.34) &3137 &64 (5.57)   &97.2  & 287 (2.95)  &2.00\\  \hline 
Yaman~\cite{yaman2021hardware} & 9508 (54.63) &2684 &16 (9.19)   &172  & 1304 (7.58)  &0.81 \\  \hline 
\textbf{Ours (Two-parallel)} &\textbf{15842 (1.58)}  &\textbf{11110 } &\textbf{84 (0.84)} &\textbf{222}  &\textbf{222  (1.00)} &\textbf{21.31 }\\  \hline
\textbf{Ours (Four-parallel)} &\textbf{33712 (1.50)}  &\textbf{24302}  & \textbf{180   (0.80)} &\textbf{222}  &\textbf{148  (0.67)} &\textbf{42.62}\\  \hline

\end{tabular}
\vspace{-1em}
\end{table*}

\section{Performance Evaluation}\label{result}

To make a fair comparison with prior works, we implement the KyberMat designs using Verilog HDL and then map them to the AC701 evaluation kit, one of the NIST-recommended Xilinx Artix-7 series FPGAs. The experimental results and comparison are presented in \figref{comp_lut_dsp} and Table~\ref{tb_mul}. The prior works~\cite{xing2021compact,guo2021efficient,bisheh2021instruction,zhao2022high,yaman2021hardware} are selected to compare with the proposed low-latency hardware design based on the same hardware platform and Kyber's parameter (i.e., $n=256$, and $q=3329$).  Two performance metrics, area and timing performances, are mainly derived in terms of LUTs (look-up tables), FFs (flip-flops), DSPs (digital signal processors), clock frequency, clock cycles, and throughput. 

\subsection{Theoretical analysis and experimental results for KyberMat accelerator in Kyber-512, Kyber-768, and Kyber-1024}
\textbf{Theoretical analysis:}  The computational complexities of different security levels, specifically Kyber-512, Kyber-768, and Kyber-1024, are primarily determined by the dimension of the matrix or vector. 

When theoretically analyzing and comparing the computational complexity for matrix-vector polynomial multiplication in NTT-domain (i.e., excluding the NTT and iNTT computation), the optimized algorithm reduces the number of modular multiplications and modular additions/subtractions. Table~\ref{tb_complexity} presents the computational complexity analysis for the matrix-vector polynomial multiplication in NTT-domain from different approaches when using the fast two-parallel structure. 

It shows that the optimized algorithm achieves an average $15.97\%$ reduction in modular multiplications and a $30.40\%$ reduction in modular additions, compared to the approach presented in \cite{xing2021compact}, when $k=\{2,3,4\}$.  Compared to the conventional method of the Kyber scheme~\cite{zhou2019preprocess}, it utilizes $70.57\%$ fewer modular additions, but the optimized algorithm reduces $33.56\%$ modular multiplications. Note that modular multiplication is much more costly than modular addition. Hence, the proposed optimization algorithm significantly reduces the overall computational complexity compared to prior designs.

\begin{table}[htbp]
\centering
\caption{Computational complexity of matrix-vector polynomial multiplication (excluding NTT/iNTT) using fast two-parallel structure }\label{tb_complexity}
    \begin{tabular}{|c||c|c|}
\hline
\multirow{2}{*}{Algorithm } & \# ModMult & \# ModAdd/Sub \\
&($k=2$, $n=256$) &($k=2$, $n=256$)\\\hline
\cite{xing2021compact} & $2k^2n$ (2048)& $\frac{7k^2n}{2}-kn$ (3072) \\
\hline
\cite{zhou2019preprocess} & $\frac{5k^2n}{2}$ (2560)& $k^2n-kn$ (512) \\
\hline
\textbf{Proposed} & \textbf{$\frac{kn+3k^2n}{2}$} (1792)& \textbf{$\frac{kn+4k^2n}{2}$} (2304) \\
\hline
\end{tabular}
\end{table}

Table \ref{tb_complexity_four} presents the computational complexity analysis for matrix-vector polynomial multiplication in the NTT-domain using the fast four-parallel structure, with and without the sub-structure sharing technique. The results indicate that the proposed technique leads to an average reduction of $22.43\%$ and $37.17\%$ in the number of modular multiplications and modular additions/subtractions, respectively, for Kyber-512, Kyber-768, and Kyber-1024 security-level ($k=\{2,3,4\}$).
\begin{table}[htbp]
\centering
\caption{Computational complexity of matrix-vector polynomial multiplication (excluding NTT/iNTT) using fast four-parallel structure}\label{tb_complexity_four}
   \scalebox{0.95}{ 
\begin{tabular}{|c||c|c|}
\hline
\multirow{2}{*}{Algorithm } & \# ModMult & \# ModAdd/Sub \\
&($k=2$, $n=256$) &($k=2$, $n=256$)\\\hline
w.o. Sub-struc. Share & $\frac{13k^2n}{4}$ (3328) & $\frac{(38k^2- 9k)n}{4}$  (8576) \\
\hline
\textbf{w. Sub-struc. Share}& $\frac{9k^2n+3kn}{4}$ (2688) & $(4k^2+4k)n$  (6144)\\
\hline
\end{tabular}
}
\end{table}

\textbf{FPGA results:} \figref{comp_lut_dsp} shows the FPGA implementation results for our KyberMat accelerator in Kyber-512, Kyber-768, and Kyber-1024. The area consumption and clock frequency for one matrix-vector polynomial multiplication in the NTT-domain module (i.e., excluding the NTT and iNTT computation modules) are separately presented in Table~\ref{tb_matrix_vector} as well. In a cryptosystem, the speed of the Encaps (encryption) and Decaps (decryption) processes plays a vital role in determining the usability of the overall application. Recognizing this critical metric, we prioritize the timing performance in the designs, distinguishing our approach from previous compact architecture designs that rely on limited hardware resources. We adopt a trade-off strategy that involves dedicating more hardware resources to achieve higher clock frequency, lower clock cycles, and higher throughput.

\textbf{Latency and speed analysis:} For the timing performance, the proposed design maintains nearly constant clock cycle consumption when the security level grows. Since more hardware resources are devoted, the latency in terms of the clock cycle is reduced, which can be summarized as
\begin{equation}
    T_{Lat} = \frac{n}{L} - 2 + N_{pipe},
\end{equation}
where $L$ is the level of parallelism ($L=2$ when using the fast two-parallel structure), and $N_{pipe}$ represents the additional clock cycles in pipelining stages added to the data-path in order to reduce the critical path. Note that the latency is considered as the number of clock cycles elapsed between the first data in and the last data out. In the proposed implementation, each modular multiplier is pipelined by five stages (i.e., $N_{pipe} = 5$). 
After employing additional pipelining stages into the data-paths, the critical path only requires 4.4 $ns$ among Kyber-512, Kyber-768, and Kyber-1024.
\begin{figure}[htbp]
\centering
\subfigure[DSP usage \textit{versus} clock cycle]
{\resizebox{0.22\textwidth}{!}{%
\includegraphics[]{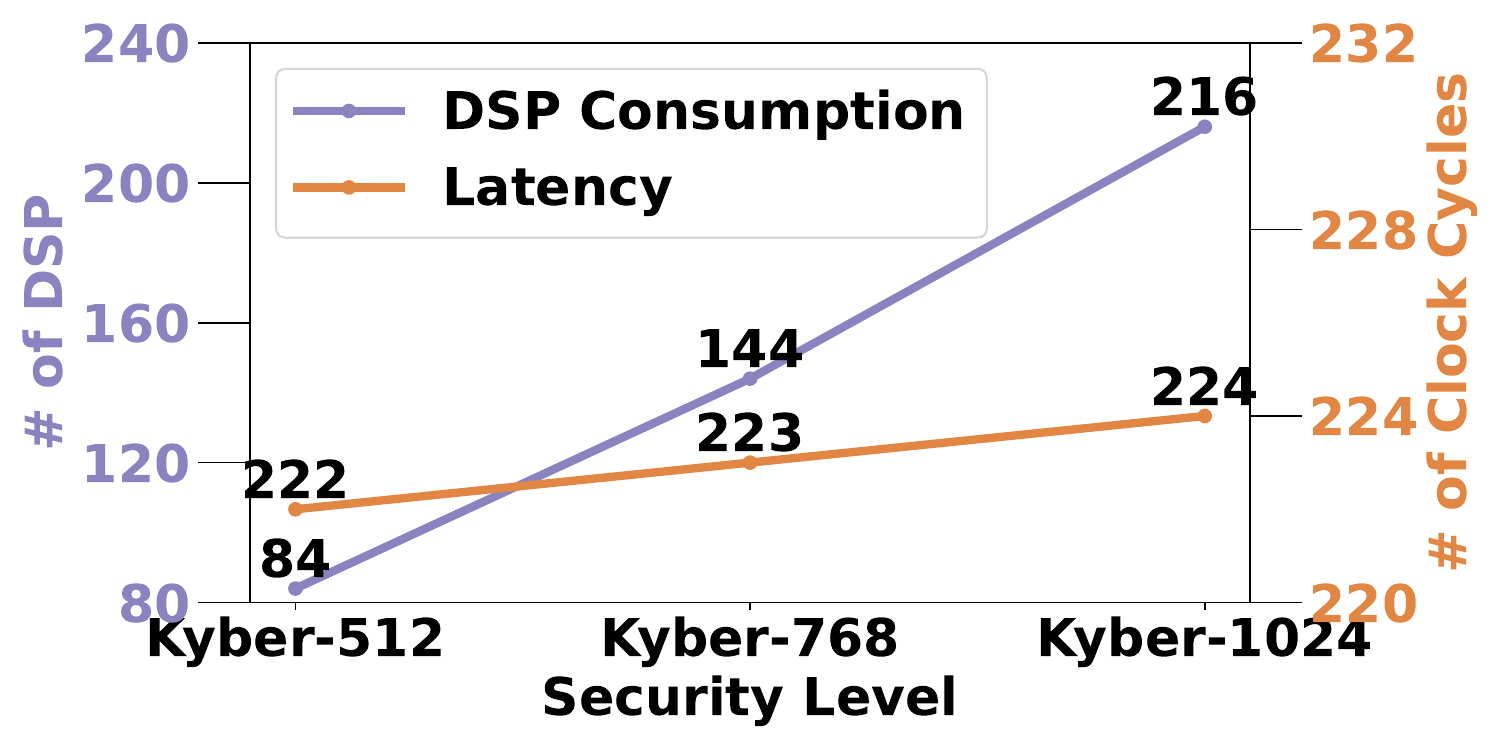}}
}
\subfigure[LUT usage \textit{versus} clock cycle]{
\resizebox{0.22\textwidth}{!}{%
\includegraphics[]{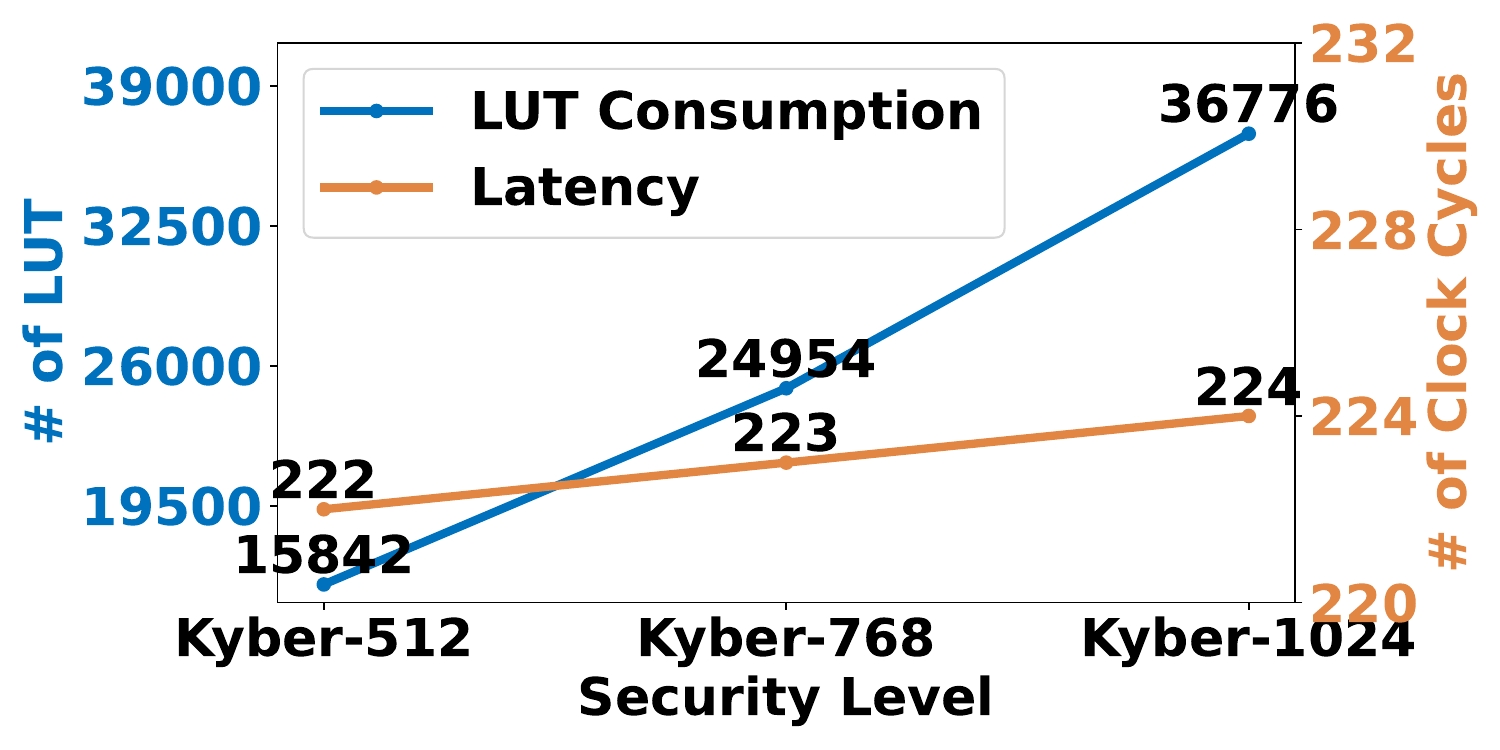}}
}
\vspace{-1em}
\caption{\small{Artix-7 FPGA implementation result for KyberMat accelerator using two-parallel structure based on different security levels in 222 MHz clock frequency.}} 
\figlbl{comp_lut_dsp}
\vspace{-0.5em}
\end{figure}

\textbf{Observation and analysis:} Despite utilizing higher numbers of LUTs, DSPs, and FFs in our proposed designs, the growth trend of LUTs/DSPs usage follows a linear trajectory, with the overhead in the number of LUTs or DSPs only increasing by a factor of around $1.59\times$ upon rising to the next higher security level. This is mainly due to the advantages provided by sub-structure sharing. Notably, the proposed design still satisfies the hardware resource constraints of the recommended Artix-7 FPGA. For instance, the proposed low-latency design for the expensive Kyber-1024 only utilizes $9.08\%$, $4.17\%$, and $13.78\%$ over the total LUTs, FFs, and DSPs resources provided by the Artix-7 FPGA, respectively, as presented in \figref{comp_lut_dsp}.

\begin{table}[htbp]
  \centering
  \caption{Area and timing performance for matrix-vector polynomial multiplication (excluding NTT/iNTT) module in Artix-7 FPGA}\label{tb_matrix_vector}
 \scalebox{0.98} {
\begin{tabular}{|c||c|c|c|c|c|}
\hline
Security-level  &LUTs &FFs &DSPs  &Freq.[MHz] &$N_{pipe}$\\ \hline 
Kyber-512  &3526   &2996 &28   &222 &12 \\  \hline
Kyber-768 &6480   &6030 &60   &222  &13\\  \hline
Kyber-1024 &12144   &11158 &104   &222  &14\\  \hline
\end{tabular}
}
\vspace{-1em}
\end{table}

\subsection{Comparison with prior works}
We then discuss the performance of the proposed matrix-vector polynomial multiplication accelerator designs based on the two-parallel and four-parallel structures, and compare them with prior works for the case when Kyber-512 security level ($k=2$), as presented in Table~\ref{tb_mul}.

\textbf{Reduced execution time:} Regarding the timing performance, the proposed low-latency design outperforms other designs in terms of clock cycles and clock frequency, thus reducing execution time significantly when compared to prior works. 
Note that the execution time is computed as the number of clock cycles divided by frequency, denoted in microseconds ($\mu s$).

The compact designs in the literature,  which are memory-based, often suffer from a communication overhead as all intermediate results must be read from and written to memory.
For example, the compact design in~\cite{xing2021compact} requires 512, 256, and 576 clock cycles for NTT computation, point-wise multiplication, and iNTT computation, respectively, with a clock frequency of 161MHz.

In contrast, our architectures are highly optimized for pipelining, minimizing the critical path. Consequently, our low-latency design using a fast two-parallel structure reduces execution time by $89.74\%$ on average, while using a fast four-parallel structure can further reduce the average execution time by $93.16\%$.

\textbf{High-throughput:}  
In this paper, we uses the block processing time (BPT) to evaluate the performance of a real-time architecture, defined as the time to process $256\cdot k$ input coefficients or output samples. Specifically, the BPT for the low-latency design utilizing a fast two-parallel structure is 64 clock cycles. The BPT is decreased to 32 clock cycles when a fast four-parallel structure is deployed.

Each sample is 12-bit, leading to the total number of input bit-stream is $(256\cdot k\cdot 12)$ bits.
As the proposed designs operate at a clock frequency of 222MHz, the throughput in low-latency design can be calculated as $\frac{256\cdot 12\cdot k \cdot 222}{BPT\cdot 10^3}$ Gb/s  when the system is in steady-state (i.e., after the first data comes out). This is equivalent to $4k$ samples per clock cycle and $8$ samples per clock cycle for a fast two-parallel structure. The throughput of the design using a fast four-parallel structure is doubled since 16 data-paths are placed in parallel. 

However, since the PEs in the prior memory-based designs have to be reconfigured to execute different operations, no data from the new input sequence can be loaded in before the entire matrix-vector polynomial multiplication computation is finished, which thus results in low throughput. 
As illustrated in Table~\ref{tb_mul}, our low-latency design using the fast two-parallel structure improves throughput by $65.81$ times compared to the prior designs, while the improvement enhances to $131.63$ times when using the fast four-parallel structure.

Furthermore, the low-latency designs using the fast two-parallel and four-parallel structures outperform the previous works in terms of throughput per DSP (TPD) and throughput per LUT (TPL). The results demonstrate an improvement of around $94.05\%$ and $87.34\%$ in TPL performance, and $87.63\%$ and $73.50\%$ in TPD performance, respectively, for two- and four-parallel designs.

\textbf{Hardware cost efficiency analysis:} The proposed designs demand more LUTs, FFs, and DSPs in trading off for speed.  For instance, the proposed low-latency design using the fast two-parallel structure requires around $55.28\%$ and $78.10\%$ more LUTs and DSPs than prior works. The LUTs and DSPs consumption overhead increases to $76.75\%$ and $89.78\%$, respectively, when using the fast four-parallel structure. 
To make a fair comparison between the prior compact architectures and the proposed designs, this paper also considers the Area Timing Square Product (AT$^2$P) to jointly evaluate area performance and timing performance, as speed is more important in the proposed design. 
The AT$^2$P results with respect to the DSP and LUT usages presented in Table~\ref{tb_mul} further demonstrate the superiority of the proposed designs over the previous works.

\section{Conclusion}\label{conclusion}
This paper proposes a novel efficient low-latency matrix-vector polynomial multiplication algorithm for the Kyber PQC scheme to reduce the number of modular multiplications and additions required. The FPGA experimental results demonstrate that the proposed designs achieve a better timing performance compared to the prior works. Although two-parallel and four-parallel structures are considered in this paper, other parallelism levels and other structures can be incorporated depending on application requirements.
\section*{ACKNOWLEDGEMENT}
This work is supported in part by the Semiconductor Research Corporation under contract number 2020-HW-2998,
and the NSF under Grant numbers CCF-2243052 and CCF-2243053.

\bibliographystyle{IEEEtran}
\bibliography{main}

\begin{thebibliography}{10}
\providecommand{\url}[1]{#1}
\csname url@samestyle\endcsname
\providecommand{\newblock}{\relax}
\providecommand{\bibinfo}[2]{#2}
\providecommand{\BIBentrySTDinterwordspacing}{\spaceskip=0pt\relax}
\providecommand{\BIBentryALTinterwordstretchfactor}{4}
\providecommand{\BIBentryALTinterwordspacing}{\spaceskip=\fontdimen2\font plus
\BIBentryALTinterwordstretchfactor\fontdimen3\font minus
  \fontdimen4\font\relax}
\providecommand{\BIBforeignlanguage}[2]{{%
\expandafter\ifx\csname l@#1\endcsname\relax
\typeout{** WARNING: IEEEtran.bst: No hyphenation pattern has been}%
\typeout{** loaded for the language `#1'. Using the pattern for}%
\typeout{** the default language instead.}%
\else
\language=\csname l@#1\endcsname
\fi
#2}}
\providecommand{\BIBdecl}{\relax}
\BIBdecl

\bibitem{bos2018crystals}
R.~Avanzi, J.~Bos, L.~Ducas, E.~Kiltz, T.~Lepoint, V.~Lyubashevsky, J.~M.
  Schanck, P.~Schwabe, G.~Seiler, , and D.~Stehl{\'e}, ``{CRYSTALS}--kyber:
  Algorithm specification and supporting documentation (version 3.02),''
  Round-3 submission to the NIST Post-Quantum Cryptography Standardization
  Project, 2020, \url{https://cryptojedi.org/papers/\#kybernistr3}.

\bibitem{regev2009lattices}
O.~Regev, ``On lattices, learning with errors, random linear codes, and
  cryptography,'' \emph{Journal of the ACM (JACM)}, vol.~56, no.~6, pp. 1--40,
  2009.

\bibitem{langlois2015worst}
A.~Langlois and D.~Stehl{\'e}, ``Worst-case to average-case reductions for
  module lattices,'' \emph{Designs, Codes and Cryptography}, vol.~75, no.~3,
  pp. 565--599, 2015.

\bibitem{potkonjak1996multiple}
M.~Potkonjak, M.~B. Srivastava, and A.~P. Chandrakasan, ``Multiple constant
  multiplications: Efficient and versatile framework and algorithms for
  exploring common subexpression elimination,'' \emph{IEEE Transactions on
  Computer-Aided Design of Integrated Circuits and Systems}, vol.~15, no.~2,
  pp. 151--165, 1996.

\bibitem{parhi2007vlsi}
K.~K. Parhi, \emph{{VLSI} digital signal processing systems: design and
  implementation}.\hskip 1em plus 0.5em minus 0.4em\relax John Wiley \& Sons,
  1999.

\bibitem{parker1997low}
D.~A. Parker and K.~K. Parhi, ``Low-area/power parallel {FIR} digital filter
  implementations,'' \emph{Journal of VLSI signal processing systems for
  signal, image and video technology}, vol.~17, no.~1, pp. 75--92, 1997.

\bibitem{cheng2004hardware}
C.~Cheng and K.~K. Parhi, ``Hardware efficient fast parallel {FIR} filter
  structures based on iterated short convolution,'' \emph{IEEE Transactions on
  Circuits and Systems I: Regular Papers}, vol.~51, no.~8, pp. 1492--1500,
  2004.

\bibitem{zhou2019preprocess}
S.~Zhou, H.~Xue, D.~Zhang, K.~Wang, X.~Lu, B.~Li, and J.~He,
  ``Preprocess-then-{NTT} technique and its applications to {Kyber} and {New
  Hope},'' in \emph{Information Security and Cryptology: 14th International
  Conference, Inscrypt 2018, Fuzhou, China, December 14-17, 2018, Revised
  Selected Papers 14}.\hskip 1em plus 0.5em minus 0.4em\relax Springer, 2019,
  pp. 117--137.

\bibitem{xing2021compact}
Y.~Xing and S.~Li, ``A compact hardware implementation of {CCA}-secure key
  exchange mechanism {CRYSTALS-KYBER} on {FPGA},'' \emph{IACR Transactions on
  Cryptographic Hardware and Embedded Systems}, pp. 328--356, 2021.

\bibitem{zhu2021ntt}
Y.~Zhu, Z.~Liu, and Y.~Pan, ``When {NTT} meets karatsuba: preprocess-then-{NTT}
  technique revisited,'' in \emph{Information and Communications Security: 23rd
  International Conference, ICICS 2021, Chongqing, China, November 19-21, 2021,
  Proceedings, Part II}.\hskip 1em plus 0.5em minus 0.4em\relax Springer, 2021,
  pp. 249--264.

\bibitem{fujisaki1999secure}
E.~Fujisaki and T.~Okamoto, ``Secure integration of asymmetric and symmetric
  encryption schemes,'' in \emph{Annual international cryptology
  conference}.\hskip 1em plus 0.5em minus 0.4em\relax Springer, 1999, pp.
  537--554.

\bibitem{ravi2022side}
P.~Ravi, A.~Chattopadhyay, J.~P. D'Anvers, and A.~Baksi, ``Side-channel and
  fault-injection attacks over lattice-based post-quantum schemes {(Kyber,
  Dilithium)}: Survey and new results,'' \emph{Cryptology ePrint Archive},
  2022.

\bibitem{lyubashevsky2008swifft}
V.~Lyubashevsky, D.~Micciancio, C.~Peikert, and A.~Rosen, ``{SWIFFT}: A modest
  proposal for {FFT} hashing,'' in \emph{International Workshop on Fast
  Software Encryption}.\hskip 1em plus 0.5em minus 0.4em\relax Springer, 2008,
  pp. 54--72.

\bibitem{bisheh2021instruction}
M.~Bisheh-Niasar, R.~Azarderakhsh, and M.~Mozaffari-Kermani, ``Instruction-set
  accelerated implementation of {CRYSTALS-kyber},'' \emph{IEEE Transactions on
  Circuits and Systems I: Regular Papers}, vol.~68, no.~11, pp. 4648--4659,
  2021.

\bibitem{aikata2022kali}
A.~Aikata, A.~C. Mert, M.~Imran, S.~Pagliarini, and S.~S. Roy, ``{KaLi}: A
  crystal for post-quantum security using {Kyber} and {Dilithium},'' \emph{IEEE
  Transactions on Circuits and Systems I: Regular Papers}, 2022.

\bibitem{hu2022ac}
X.~Hu, J.~Tian, M.~Li, and Z.~Wang, ``{AC-PM}: An area-efficient and
  configurable polynomial multiplier for lattice based cryptography,''
  \emph{IEEE Transactions on Circuits and Systems I: Regular Papers}, 2022.

\bibitem{lucke94}
L.~E. Lucke and K.~K. Parhi, ``Parallel processing architectures for rank order
  and stack filters,'' \emph{IEEE Transactions on Signal Processing}, vol.~42,
  no.~5, pp. 1178--1189, 1994.

\bibitem{oppenheim2009discrete}
A.~V. Oppenheim and R.~W. Schafer, \emph{Discrete-time signal
  processing}.\hskip 1em plus 0.5em minus 0.4em\relax Prentice Hall Press, USA,
  3rd edition, 2009.

\bibitem{yuan2020high}
T.~Yuan, W.~Liu, J.~Han, and F.~Lombardi, ``High performance {CNN} accelerators
  based on hardware and algorithm co-optimization,'' \emph{IEEE Transactions on
  Circuits and Systems I: Regular Papers}, vol.~68, no.~1, pp. 250--263, 2020.

\bibitem{denkinger2022vwr2a}
B.~W. Denkinger, M.~Pe{\'o}n-Quir{\'o}s, M.~Konijnenburg, D.~Atienza, and
  F.~Catthoor, ``{VWR2A}: a very-wide-register reconfigurable-array
  architecture for low-power embedded devices,'' in \emph{Proceedings of the
  59th ACM/IEEE Design Automation Conference}, 2022, pp. 895--900.

\bibitem{cheng2020fast}
C.~Cheng and K.~K. Parhi, ``Fast {2D} convolution algorithms for convolutional
  neural networks,'' \emph{IEEE Transactions on Circuits and Systems I: Regular
  Papers}, vol.~67, no.~5, pp. 1678--1691, 2020.

\bibitem{tan2023high}
W.~Tan, A.~Wang, X.~Zhang, Y.~Lao, and K.~K. Parhi, ``High-speed {VLSI}
  architectures for modular polynomial multiplication via fast filtering and
  applications to lattice-based cryptography,'' \emph{IEEE Transactions on
  Computers}, vol.~72, no.~9, pp. 2454--2466, 2023.

\bibitem{xing2019efficient}
Y.~Xing and S.~Li, ``An efficient implementation of the {NewHope-Simple} key
  exchange on {FPGA}s,'' \emph{IEEE Transactions on Circuits and Systems I:
  Regular Papers}, vol.~67, no.~3, pp. 866--878, 2019.

\bibitem{li2022reconfigurable}
M.~Li, J.~Tian, X.~Hu, and Z.~Wang, ``Reconfigurable and high-efficiency
  polynomial multiplication accelerator for {CRYSTALS-Kyber},'' \emph{IEEE
  Transactions on Computer-Aided Design of Integrated Circuits and Systems},
  2022.

\bibitem{mou1991short}
Z.-J. Mou and P.~Duhamel, ``Short-length {FIR} filters and their use in fast
  nonrecursive filtering,'' \emph{IEEE Transactions on Signal Processing},
  vol.~39, no.~6, pp. 1322--1332, 1991.

\bibitem{tan2023parentt}
W.~Tan, S.-W. Chiu, A.~Wang, Y.~Lao, and K.~K. Parhi, ``{PaReNTT}: Low-latency
  parallel residue number system and {NTT}-based long polynomial modular
  multiplication for homomorphic encryption,'' \emph{arXiv preprint
  arXiv:2303.02237}, 2023.

\bibitem{zhao2022high}
Y.~Zhao, R.~Xie, G.~Xin, and J.~Han, ``A high-performance domain-specific
  processor with matrix extension of {RISC-V} for {Module-LWE} applications,''
  \emph{IEEE Transactions on Circuits and Systems I: Regular Papers}, vol.~69,
  no.~7, pp. 2871--2884, 2022.

\bibitem{nejatollahi2020exploring}
H.~Nejatollahi, S.~Shahhosseini, R.~Cammarota, and N.~Dutt, ``Exploring energy
  efficient quantum-resistant signal processing using array processors,'' in
  \emph{ICASSP 2020 IEEE International Conference on Acoustics, Speech and
  Signal Processing (ICASSP)}.\hskip 1em plus 0.5em minus 0.4em\relax IEEE,
  2020, pp. 1539--1543.

\bibitem{hirner2023proteus}
F.~Hirner, A.~C. Mert, and S.~S. Roy, ``{PROTEUS}: A tool to generate pipelined
  number theoretic transform architectures for {FHE} and {ZKP} applications,''
  \emph{Cryptology ePrint Archive}, 2023.

\bibitem{ayinala2011pipelined}
M.~Ayinala, M.~Brown, and K.~K. Parhi, ``Pipelined parallel {FFT} architectures
  via folding transformation,'' \emph{IEEE Transactions on Very Large Scale
  Integration Systems}, vol.~20, no.~6, pp. 1068--1081, 2012.

\bibitem{guo2021efficient}
W.~Guo, S.~Li, and L.~Kong, ``An efficient implementation of {KYBER},''
  \emph{IEEE Transactions on Circuits and Systems II: Express Briefs}, vol.~69,
  no.~3, pp. 1562--1566, 2022.

\bibitem{yaman2021hardware}
F.~Yaman, A.~C. Mert, E.~{\"O}zt{\"u}rk, and E.~Sava{\c{s}}, ``A hardware
  accelerator for polynomial multiplication operation of {CRYSTALS-KYBER} {PQC}
  scheme,'' in \emph{2021 Design, Automation \& Test in Europe Conference \&
  Exhibition (DATE)}.\hskip 1em plus 0.5em minus 0.4em\relax IEEE, 2021, pp.
  1020--1025.

\end{thebibliography}

\end{document}